\UseRawInputEncoding
\documentclass[superscriptaddress, twocolumn, amsmath, amssymb, aps,pre,letterdraft, notitlepage,longbibliography, 10pt]{revtex4-1} 
\usepackage{graphicx,graphics,epsfig,subfigure,times,bm,bbm,amssymb,amsmath,amsfonts,amsthm,mathrsfs,MnSymbol}
\usepackage[matrix,frame,arrow]{xypic}
\usepackage[normalem]{ulem}
\usepackage{slashed}
\usepackage{dcolumn}
\usepackage{tabularx}
\usepackage{color}
\usepackage[usenames,dvipsnames,svgnames,table]{xcolor}
\usepackage[english]{babel}
\usepackage{verbatim}
\definecolor{darkblue}{rgb}{0.0,0.0,0.3}
\usepackage[colorlinks=true,
            linkcolor=red,
            urlcolor= darkblue,
            citecolor=blue]{hyperref}

\newcommand{\bea}{\begin{eqnarray}}
\newcommand{\eea}{\end{eqnarray}}

\begin{document}
\title{Coherences and the thermodynamic uncertainty relation: 
Insights from quantum absorption refrigerators}

\author{Junjie Liu}
\address{Department of Chemistry and Centre for Quantum Information and Quantum Control,
University of Toronto, 80 Saint George St., Toronto, Ontario, M5S 3H6, Canada}
\author{Dvira Segal}
\address{Department of Chemistry and Centre for Quantum Information and Quantum Control,
University of Toronto, 80 Saint George St., Toronto, Ontario, M5S 3H6, Canada}
\address{Department of Physics, 60 Saint George St., University of Toronto, Toronto, Ontario, Canada M5S 1A7}

\begin{abstract}
The thermodynamic uncertainty relation, originally derived for classical Markov-jump processes, provides a trade-off relation
between precision and dissipation, deepening our understanding of the performance of quantum thermal machines. 
Here, we examine the interplay of quantum system coherences and heat current
fluctuations on the validity of the thermodynamics uncertainty relation 
in the quantum regime. To achieve the current statistics, we perform a full counting statistics simulation
of the Redfield quantum master equation. 
We focus on steady-state quantum absorption refrigerators where nonzero coherence between eigenstates can either 
suppress or enhance the cooling power, compared with the incoherent limit. In either
scenario, we find enhanced relative noise of the cooling power 
(standard deviation of the power over the mean)  
in the presence of system coherence, thereby corroborating the thermodynamic uncertainty relation. 
Our results indicate that fluctuations necessitate consideration when assessing the 
performance of quantum coherent thermal machines.
\end{abstract}

\date{\today}

\maketitle

\section{Introduction}
Quantum thermodynamics is an emerging research field concerning the thermodynamics and 
nonequilibrium statistical mechanics of open quantum nanoscale systems with the full inclusion of quantum effects \cite{Scovil.59.PRL,Geva.92.JCP,Gemmer.09.NULL,Seifert.12.RPP,Pekola.15.NP,Kosloff.15.E,Goold.16.JPA,Anders.16.CP,Binder.18.NULL,Seifert.19.ARCMP,Pekola.19.ARCMP}. In the quantum realm, basic notions such as heat and work of classical thermodynamics need to be reexamined and refined (see for example, Refs. \cite{Talker.07.PRE,Talkner.16.PRE,Talkner.20.RMP}), leading to, for instance, intriguing microscopic unravelling of the second law of thermodynamics \cite{Esposito.09.RMP,Campisi.11.RMP}.

A tantalizing prospect of the field of quantum thermodynamics is to devise and realize quantum thermal machines (QTMs) that transform heat to useful work or use work to refrigerate \cite{Scovil.59.PRL,Alicki.79.JPA,Geva.92.JCPa,Giazotto.06.RMP,Filliger.07.PRL,Esposito.10.PRLa,Linden.10.PRL,Seifert.12.RPP,Levy.12.PRL,Abah.12.PRL,Blickle.12.NP,Kosloff.14.ARPC,Rossnagel.14.PRL,Thierschmann.15.NN,Dechant.15.PRL,Harris.16.PRA,Robnagel.16.S,Uzdin.15.PRX,Hofer.16.PRB,Hofer.16.PRBa,Zou.17.PRL,Martinez.15.NP,Ghosh.17.PNAS,Klaers.17.PRX,Elouard.17.PRL,Abah.17.EPL,Benenti.17.PR,Ronzani.18.NP,Josefsson.18.NN,Friedman.18.NJP,Kilgour.18.PRE,Elouard.18.PRL,Pietzonka.19.PRX,Assis.19.PRL,Mitchison.19.CP,Peterson.19.PRL,Klatzow.19.PRL,Lindenfels.19.PRL,Buffoni.19.PRL,Abiuso.20.PRL,Carollo.20.PRL,Hartmann.20.PRR,Horne.20.NPJQI,Ono.20.PRL,Bhandari.20.PRB}. 
Efficient QTMs, capable of exploiting genuine quantum effects and outperforming their classical counterparts, are of paramount importance for future quantum technologies. In this regard, the discussion of whether quantum coherence can be an advantageous resource in the operation of QTMs is an ongoing and vivid topic of this field \cite{Scully.S.03,Zhang.07.PRA,Dillenschneider.09.EPL,Scully.11.PNAS,Dorfman.13.PNAS,Rahav.12.PRA,Killoran.15.JCP,Leggio.15.PRA,Niedenzu.15.PRE,Gelbwaser.15.SR,Brandner.15.NJP,Mitchison.15.NJP,Uzdin.15.PRX,Su.16.PRE,Turkpence.16.PRE,Uzdin.16.PRA,Mehta.17.PRE,Chen.17.E,Holubec.18.JLTP,Kilgour.18.PRE,Wertnik.18.JCP,Holubec.19.JCP,Latune.19.SR,Latune.20.NULL,Francica.20.PRL}. 

At the nanoscale, fluctuations become significant, 
implying that one should inspect nonequilibrium fluctuations of thermodynamic quantities for characterizing the performance of QTMs \cite{Verley.14.NC,Seifert.12.RPP,Rahav.12.PRA,Esposito.15.PRBa,Bijay.15.PRB,Polettini.15.PRL,Jiang.15.PRL,Campisi.15.NJP,Campisi.16.NC,Holubec.17.PRE,Segal.18.PRE,HavaC,Friedman.18.NJP,Holubec.18.PRL,Manikandan.19.PRL,Vroylandt.20.PRL}. Recently, a conceptual advance, termed thermodynamic uncertainty relation (TUR) \cite{Barato.15.PRL,Gingrich.16.PRL,Horowitz.19.NP}, is offering new insights into the characteristics of steady-state QTMs in terms of a trade-off relation between power fluctuation and efficiency \cite{Pietzonka.18.PRL}. While the original TUR was derived for systems described by classical Markov jump processes, quantum generalizations of TURs have been conceived for steady-state \cite{Guarnieri.19.PRR} and cyclic \cite{Miller.20.A} QTMs.
To assess the performance of QTMs, 
it is necessary to explore how quantum effects such as coherence contribute to the relative noise
and to the behavior of the TUR \cite{Ptaszynski.18.PRB,Bijay.18.PRB,Liu.19.PRE,Bret.20.A,Cangemi.20.PRB}. 

Here, we focus on models for a quantum absorption refrigerator (QAR), 
a steady-state QTM that continuously pumps heat from a cold bath into a hot bath by consuming power from a (very hot) `work' bath. The study of  QAR developed from early studies of a three-level maser, an engine \cite{Scovil.59.PRL}. 
With rapid developments in quantum thermodynamics, recent years have witnessed a great number of investigations on various aspects of QARs \cite{Linden.10.PRL,Brunner.12.PRE,Levy.12.PRL,Correa.13.PRE,Kosloff.14.ARPC,Correa.14.PRE,Correa.15.PRE,Hofer.16.PRBa,Mitchison.16.QST,Mu.17.NJP,Gonzalez.17.NJP,Segal.18.PRE,Correa.14.SR,Mitchison.19.CP}. Despite significant progress on the subject, the interplay of coherence and fluctuation in the performance of QARs remains largely unexplored, although studies have revealed the pros and cons of system coherence on cooling power \cite{Kilgour.18.PRE,Holubec.18.JLTP,Mitchison.19.CP}. 

Focusing on QARs in which the system coherence  
was shown to have a nontrivial effect on the cooling power \cite{Kilgour.18.PRE,Holubec.18.JLTP,Holubec.19.JCP}, 
the objectives of the present study are twofold: 
(i) We aim to identify the role of steady state system coherence (between system energy eigenstates) 
 on power fluctuations. 
If system coherences are systematically linked to 
reduced fluctuations, they become a useful resource for QARs if the net power is instead enhanced. 
Conversely, if fluctuations consistently increase when system coherences exist  (compared to the incoherent limit), 
then power boost associated with coherences may not be profitable.
(ii) We aim to test the behavior of the TUR ratio (relative noise times entropy production)  
\cite{Barato.15.PRL,Gingrich.16.PRL,Guarnieri.19.PRR} in the presence of system coherences. 
Can coherences improve this trade-off relation, 
i.e. allow the system to operate closer to the bound relative to the incoherent scenario?

To attain the heat current and its noise, 
we combine a full counting statistics formalism  \cite{Levitov.93.JETP,Levitov.96.JMP} 
within the Redfield master equation (RME) \cite{Redfield.57.IBM,Redfield.65.AMOR}.
The method is perturbative in the system-bath coupling, 
but allows arbitrarily strong internal system's coherences.

We confirm the thermodynamic consistency of the method by validating the fluctuation symmetry for heat transfer \cite{Esposito.09.RMP,Campisi.11.RMP}.
To demonstrate the nontrivial role of system coherences on the cooling current and its fluctuations, 
we compare results under the full Redfield calculations 
with those obtained in the incoherent limit where coherences vanish. 
We find that system coherences always enhance the relative noise of the cooling power 
compared with the incoherent limit, despite that it can either suppress or enhance the cooling power itself. 
Particularly, considering a model
for a QAR where system coherences enhance the cooling power \cite{Holubec.18.JLTP}, we show that power fluctuations get amplified as well, indicating that system coherences are not helpful in terms of constancy \cite{Pietzonka.18.PRL}.

As a result of the enhanced relative noise, the TUR is always satisfied by the QARs under investigation here---regardless of the presence of system coherences. Our results thus suggest that violations of TUR observed in steady-state QTMs at weak couplings 
\cite{Ptaszynski.18.PRB,Bijay.18.PRB,Liu.19.PRE} should not be attributed to nonzero system coherences. 

The paper is organized as follows. In Sec. \ref{sec:2}, we first introduce the general setup and the counting-field dressed Redfield master equation. We then briefly mention how to get currents and fluctuations from the cumulant generating function and demonstrate the thermodynamic consistency of the Redfield master equation by validating the fluctuation symmetry for a V-shaped system. In Sec. \ref{sec:3}, we focus on two concrete examples of steady-state QARs and provide detailed simulation results for the cooling current, its fluctuations, and the TUR. We conclude in Sec. \ref{sec:4}.

\section{Model and methodology}
\label{sec:2}

In this section, we first introduce  the general modeling of QTMs as open quantum systems connected to multiple heat baths. 
We then lay out the counting-field dressed Redfield master equation ($\chi-$ RME) for the sake of completeness and 
present expressions for currents and fluctuations from the cumulant generating function. 
Finally, we check the thermodynamic consistency of the counting-field dressed Redfield master equation by verifying the fluctuation symmetry of heat exchange for the V-shaped system. The behavior of the heat current and coherences in the V-shaped system closely relate to the cooling characteristics of our model I [see Fig. \ref{fig:fig3} (a)] for a QAR \cite{Kilgour.18.PRE}. 
As such, verifying the fluctuation symmetry of heat exchange for the V-shaped system indicates on the corresponding behavior for QARs.

\subsection{Hamiltonian for QTMs}
QTMs can be modeled as open quantum systems consisting of a central system ($s$) coupled to multiple (counted by $v$) bosonic heat baths ($b$). The total system-bath Hamiltonian reads (setting $\hbar=1$ and $k_B=1$ hereafter)
\begin{equation}\label{eq:h_tot}
\hat H~=~\hat H_s+\sum_v\hat H_b^v+\hat H_{sb},~~~\hat H_{sb}~=~\sum_v\hat S_v\otimes \hat B_v.
\end{equation}
Here and in what follows, we imply the tensor product with the identity so that $\hat H_s$ will be used instead of $\hat H_s\otimes \mathbb{I}_b$ with $\mathbb{I}_b$ an identity matrix in the bath subspace and so on. The working substance $\hat H_s$ constitutes a few-level quantum system. The bosonic heat baths are assumed to be harmonic,
\begin{equation}
\hat H_b^v~=~\sum_k\omega_{kv}\hat b_{kv}^{\dagger}\hat b_{kv},
\end{equation}
with $\hat b_{kv}^{\dagger}$ ($\hat b_{kv}$) creating (annihilating) a harmonic mode $k$ with frequency $\omega_{kv}$ in $v$ bath. Here, we assume that each bath is described by a thermal equilibrium state characterized by a temperature $T_v$. $\hat S_v$ and $\hat B_v$ are system and reservoir operators that form the coupling between the system and $v$ bath, respectively. We consider bilinear system-bath interactions and take $\hat B_v$ to be the displacement operators,
\begin{equation}
\hat B_v~=~\sum_k\lambda_{kv}(\hat b_{kv}^{\dagger}+\hat b_{kv})
\end{equation}
with $\lambda_{kv}$ characterizing the coupling strength between the system and the $v$ reservoir.

\subsection{Counting-field dressed Redfield master equation}
\label{sec:2R}

To study the thermodynamics of the generic QTMs defined above, we combine the full counting statistic formalism \cite{Levitov.93.JETP,Levitov.96.JMP} with the Redfield master equation \cite{Redfield.57.IBM,Redfield.65.AMOR}. 
This formalism was recently described in details in Ref. \cite{Friedman.18.NJP}.
To do so, we assign each bath a counting field $\chi_v$.
The moment generating function is defined with the two-time measurement protocol as \cite{Esposito.09.RMP}
\bea
\mathcal{Z}(\{\chi_v\},t)~\equiv~{\rm Tr} 
\left[ e^{i\sum_v\chi_v\hat H_b^v(0)}e^{-i\sum_v\chi_v\hat H_b^v(t)} \hat \rho(0) \right].
\eea
Here, operators are written in the Heisenberg picture. 
$\hat \rho(0)$ denotes the initial factorized density matrix of the system (s) and bath (b), 
$\hat \rho(0)=\hat \rho_s(0)\otimes\hat \rho_b(0)$; $\hat \rho_b(0)=\prod_{v}\exp(-\beta_v\hat H_v)/Z_v$ with $\beta_v=T_v^{-1}$ and $Z_v$ the inverse temperature and partition function for the $v$ bath, respectively. 
After some simple manipulations, we arrive at \cite{Friedman.18.NJP}
\begin{equation}
\mathcal{Z}(\{\chi_v\},t)~=~\mathrm{Tr}[\hat \rho^{\chi}(t)],
\end{equation}
where the counting-field dressed total density matrix reads
\begin{equation}
\hat \rho^{\chi}(t)~\equiv~\hat U^{-\chi}(t)\hat \rho(0)\hat U^{\chi,\dagger}(t);~~\hat U^{-\chi}(t)~\equiv~e^{-i\hat H^{-\chi}t}
\label{eq:rhox}
\end{equation}
%
with $\hat H^{-\chi}\equiv e^{-i\sum_v\chi_v\hat H_b^v/2}\hat He^{i\sum_v\chi_v\hat H_b^v/2}$ denoting the counting-field dressed total Hamiltonian. 
Notice that we have $[\hat \rho^{\chi}]^{\dagger}=\hat \rho^{-\chi}$.
Equation (\ref{eq:rhox}) can be written as a differential equation, generalizing the Liouville Equation
 for the density matrix (in the Schr\"odinger picture),
\bea
\frac{d\hat\rho^{\chi}}{dt}  = -i\hat H^{-\chi}\hat \rho^{\chi}(t) + i  \rho^{\chi}(t) \hat H^{\chi}.
\label{eq:EOM}
\eea
The moment generating function is obtained by solving this equation of motion,  then tracing $\hat \rho^{\chi}$.

Using the explicit form given by Eq.  (\ref{eq:h_tot}), we get 
\begin{equation}
\hat H^{-\chi}~=~\hat H_s+\sum_v\hat H_b^v+\sum_v\hat S_v\otimes \hat B_v^{-\chi_v},
\end{equation}
where $\hat B_v^{-\chi_v}=e^{-i\chi_v\hat H_b^v/2}\hat B_ve^{i\chi_v\hat H_b^v/2}=\sum_k\lambda_{kv}(e^{-i\chi_v\omega_{kv}/2}\hat b_{kv}^{\dagger}+\mathrm{H.c.})$ with `H.c.' denoting Hermitian conjugate hereafter.
Proceeding from Eq. (\ref{eq:EOM}) in the interaction representation,  
treating the counting-field dressed system-bath coupling as a perturbation, and then transforming back to the Schr\"odinger picture, 
 the reduced density matrix dynamics 
$\rho_{s}^{\chi}(t)\equiv {\rm Tr}_{b}[\rho^{\chi}(t)]$
can be described by the following $\chi$-RME in the energy basis $\{|n\rangle\}$ of $\hat H_s$ (detailed derivation can be found in, e.g., Ref. \cite{Friedman.18.NJP}):
\bea\label{eq:full_rme}
\frac{\partial}{\partial t}\rho_{s,nm}^{\chi}(t) &=& -i\Delta_{nm}\rho_{s,nm}^{\chi}(t)-\sum_v\sum_{jk}\Big[\mathcal{R}_{mk,kj}^{v,\ast}(\Delta_{jk})\rho_{s,nj}^{\chi}(t)\nonumber\\
&&-\mathcal{R}_{nj,km}^{\chi_v}(\Delta_{jn})\rho_{s,jk}^{\chi}(t)-\mathcal{R}_{mk,jn}^{-\chi_v^{\ast},\ast}(\Delta_{km})\rho_{s,jk}^{\chi}(t)\nonumber\\
&&+\mathcal{R}_{nj,jk}^v(\Delta_{kj})\rho_{s,km}^{\chi}(t)\Big].
\eea
Here, $\rho_{s,nm}^{\chi}(t)\equiv\langle n|\rho_s^{\chi}(t)|m\rangle$,
$\Delta_{ij}=E_i-E_j$ are energy gaps with $E_i$ eigenenergies of the subsystem in the global (eigenenergy) basis.
The superscript `$\ast$' denotes complex conjugate. 
The standard Redfield equation is recovered when counting parameters are taken to zero, 
$\mathcal{R}_{nm,lk}^v(\omega)=\left.\mathcal{R}_{nm,lk}^{\chi_v}(\omega)\right|_{\chi_v=0}$.
The transition coefficients satisfy
\bea\label{eq:trans}
\mathcal{R}_{nm,lk}^{\chi_v}(\omega) &\equiv& S_v^{nm}S_v^{lk}\int_0^{\infty}d\tau e^{i\omega\tau}\Omega_v(\chi_v+\tau),
\eea
with $\Omega_v(\tau)\equiv\langle \hat B_v(\tau)\hat B_v(0)\rangle_b$ denoting the bath correlation function evaluated using the state $\hat \rho_b(0)$. Its explicit form reads
\bea\label{eq:b_corr}
\Omega_v(\tau) 
= \int_0^{\infty}d\omega\frac{\gamma_v(\omega)}{2\pi}\Big[e^{i\omega\tau}n_B^v(\omega) 
+e^{-i\omega\tau}\Big(1+n_B^v(\omega)\Big)\Big]
\nonumber\\
\eea
with $\gamma_v(\omega)=2\pi\sum_k\lambda_{kv}^2\delta(\omega-\omega_{kv})$ and $n_B^v(\omega)$ the spectral density and Bose-Einstein distribution of $v$ bath, respectively. Without loss of generality, here we consider an Ohmic function $\gamma_v(\omega)=\alpha_v\omega e^{-\omega/\omega_c}$; $\alpha_v$ is a dimensionless system-bath coupling strength. We assume all baths have the same cutoff frequency $\omega_c$, which defines the largest energy scale in the problem. 

With Eq. (\ref{eq:b_corr}), we can evaluate the above transition coefficients 
as $\mathcal{R}_{nm,lk}^{\chi_v}(\omega)=S_v^{nm}S_v^{lk}\Gamma_{\chi_v}(\omega)$ with
\bea\label{eq:RR}
\Gamma_{\chi_v}(\omega) &\equiv& \left\{
\begin{array}{cc}
\gamma_v(\omega)e^{-i\omega\chi_v}[1+n_B^v(\omega)]/2 & \mathrm{for}~\omega>0,\\
\gamma_v(|\omega|)e^{i|\omega|\chi_v}n_B^v(|\omega|)/2 & \mathrm{for}~\omega<0.
\end{array}
\right.
\eea
Here, `$|A|$' takes the absolute value of $A$. We neglected the Lamb shifts, which is justified in the weak system-bath coupling limit.
To assess the role of nonzero system coherences, we contrast the full $\chi$-RME given by Eq. (\ref{eq:full_rme}) against its secular counterpart
in which coherences between energy eigenstates vanish \cite{Friedman.18.NJP}, 
\bea\label{eq:secular_rme}
\frac{\partial}{\partial t}\rho_{s,nn}^{\chi}(t) &=& -\sum_v\sum_{k}\Big[2\mathrm{Re}\left(\mathcal{R}_{nk,kn}^{v}(\Delta_{nk})\right)\rho_{s,nn}^{\chi}(t)\nonumber\\
&&-\mathcal{R}_{nk,kn}^{\chi_v}(\Delta_{kn})\rho_{s,kk}^{\chi}(t)\nonumber\\
&&-\mathcal{R}_{nk,kn}^{-\chi_v^{\ast},\ast}(\Delta_{kn})\rho_{s,kk}^{\chi}(t)\Big].
\eea
Hereafter, we refer to Eq. (\ref{eq:secular_rme}) as the secular $\chi$-RME with the understanding that it  represents the incoherent limit of the full $\chi$-RME. 
Below, we use ``secular limit" and ``incoherent limit" interchangeably. 

To obtain the current and its higher-order cumulants, we recast the $\chi$-RME in the Liouville space as
\begin{equation}
\frac{\partial}{\partial t}|\rho^{\chi}_s(t)\rangle\rangle~=~-\mathbb{L}_{\chi}|\rho_s^{\chi}(t)\rangle\rangle,
\end{equation}
where $|\rho_s^{\chi}\rangle\rangle\equiv(\rho_{s,11}^{\chi},\rho_{s,12}^{\chi},\cdots,\rho_{s,nm}^{\chi},\cdots,\rho_{s,NN}^{\chi})^{T}$ denotes an $N^2\times 1$ vector with $N$ the dimension of the system Hilbert space. $\mathbb{L}_{\chi}$ is an $N^2\times N^2$ matrix representing the $\chi$-dependent Liouvillian superoperator (noting that $\mathbb{L}_{\chi}$ reduces to an $N\times N$ matrix in the secular limit).  
In the steady state limit, the cumulant generating function (CGF) $G(\chi)~=~\lim_{t\to\infty}\ln\mathcal{Z}(\chi,t)/t$ is given by \cite{Esposito.09.RMP}
\begin{equation}\label{eq:cgf}
G(\chi)~=~-\mathcal{E}_0(\chi),
\end{equation}
where $\mathcal{E}_0(\chi)$ is the ground-state energy (or the eigenvalue of the smallest real part) of the superoperator $\mathbb{L}_{\chi}$. In scenarios with multiple bath,  $\chi$ should be understood as a collection of counting fields, namely, $\chi=\{\chi_v\}$. 

We note that $\mathcal{E}_0(0)=0$, that is, without counting the smallest eigenstate is zero, corresponding to
 the steady state solution. The CGF supplies all cumulants, specifically the steady state heat current 
$\langle J_v\rangle$ out of the $v$th reservoir and its fluctuation $\langle\langle J_v^2\rangle\rangle\equiv\langle J_v^2\rangle-\langle J_v\rangle^2$,
\bea\label{eq:quantity}
\langle J_v\rangle &=& -\left.\frac{\partial \mathcal{E}_0(\chi)}{\partial (i\chi_v)}\right|_{\{\chi_v\}=0}\nonumber\\
&=& -\left.\frac{\mathcal{E}_0(\chi_v)-\mathcal{E}_0(-\chi_v)}{2(i\chi_v)}\right|_{\chi_v\to0,\{\chi_{v'\neq v}\}=0},\nonumber\\
\langle\langle J_v^2\rangle\rangle &=& -\left.\frac{\partial^2 \mathcal{E}_0(\chi)}{\partial (i\chi_v)^2}\right|_{\{\chi_v\}=0}\nonumber\\
&=& \left.\frac{\mathcal{E}_0(\chi_v)+\mathcal{E}_0(-\chi_v)}{\chi_v^2}\right|_{\chi_v\to0,\{\chi_{v'\neq v}\}=0}.
\eea
Here, $\mathcal{E}_0(-\chi)$ is the ground-state energy of $\mathbb{L}_{-\chi}$. In numerical simulations presented 
below, we adopt a real value of $\chi_v=0.001-0.005$ 
to calculate $\langle J_v\rangle$ and $\langle\langle J_v^2\rangle\rangle$ according to the second lines of the above definitions. We verified that the final results of $\langle J_v\rangle$ and $\langle\langle J_v^2\rangle\rangle$ were independent of the value of $\chi_v$ we set.

With the ability to calculate currents and fluctuations, we further study the validity of the thermodynamic uncertainty relation (TUR) \cite{Barato.15.PRL,Gingrich.16.PRL}
\begin{equation}\label{eq:tur}
\frac{\langle\langle J_v^2\rangle\rangle}{\langle J_v\rangle^2}\langle \sigma \rangle \geqslant2,
\end{equation}
with $\langle \sigma\rangle =-\sum_{v}\langle J_v\rangle\beta_v$ the total entropy production rate. 
Here we adopt the convention that $\langle J_v\rangle>0$ when flowing into the system. In what follows, 
we refer to the quantity `$\frac{\langle\langle J_v^2\rangle\rangle}{\langle J_v\rangle^2}\langle \sigma\rangle$' 
as the TUR ratio. 
Note that below we find that our models for QAR always satisfy the original TUR (\ref{eq:tur}). 
As such, we obviously satisfy the generalized---and less tight bounds as discussed in Refs. \cite{Guarnieri.19.PRR,Horowitz.19.NP}.

In sum, in our procedure we construct the Liouvillians $\mathbb{L}_{\chi}$ for both the full 
Redfield and the secular Redfield equations and find their smallest eigenvalues, the CGF. 
We obtain the steady state heat current from bath $v$ and the current noise
by numerically calculating the first and second derivatives of the CGF, respectively, 
taken with respect to the counting parameter of bath $v$.
   
\subsection{Verifying the fluctuation symmetry with the $\chi$-RME}
\label{sec:2S}

Numerous studies in past years had examined the regime of validity and 
accuracy of second-order Markovian quantum master equations,
specifically comparing the Redfield master equation to the local Lindblad  master equation (LLME), which is performed in the site basis, and the eigenbasis Lindblad  master equation (ELME) 
\cite{Gemmer07,Levy14, Manas, Hofer17, Luis17, Landi18, Zambrini19}, which is performed in the global basis. 
The Redfield equation reduces to the ELME after making the secular approximation. 
The LLME is derived in the site basis, 
and it is known to miss proper thermalization further showing incorrect transport properties \cite{Levy14}. 
Comparing the predictions of the RME to exact results (when available), 
it has been generally concluded that the RME is superior over both the ELME and the LLME
\cite{Gemmer07,Manas,Luis17}.

On the other hand, the Redfield dissipator does not necessarily satisfy the 
condition of complete positivity (unlike the Lindblad dissipator).
How does this deficiency impact thermodynamical properties?
To the best of our knowledge, there are no reports on 
the impact of deviations from complete positivity on {\it steady state behavior}.  
However, for a driven system
it was shown in Ref. \cite{Argentieri} that the departure from complete positivity lead
to the violation of the second law of thermodynamics, 
with a negative entropy production at intermediate times. 

For a system coupled to three baths ($h, w, c$) with no internal leaks, the cumulant generating function
satisfies the steady state exchange fluctuation symmetry (SSFS) 
\cite{Esposito.09.RMP,Campisi.11.RMP,Segal.18.PRE,Andrieux.09.NJP,Liu.18.JCP},
\begin{equation}\label{eq:efsg}
G(\chi_c,\chi_w)~=~G(i(\beta_h-\beta_c)-\chi_c, i(\beta_h-\beta_w)-\chi_w).
\end{equation}
A fundamental question that is still not settled is whether the full
$\chi$-RME satisfies this relation, thus is consistent with nonequilibrium thermodynamics.
The secular limit of the $\chi$-RME, 
that is the eigenbasis Lindblad  master equation,  satisfies the SSFS for QAR \cite{Segal.18.PRE}. 
On the other hand, if the fluctuations symmetry is not satisfied in the full $\chi$-RME, 
it is important to find out what is the impact of this deviation on the accuracy of calculated
cumulants of transport. In other words, whether this deviation influences the small $\chi$ behavior.
This point is critical to our analysis since we are using the $\chi$-RME
to calculate the current and its fluctuations under the influence of quantum coherences.

%
\begin{figure}[tbh!]
 \centering
\includegraphics[width=1\columnwidth] {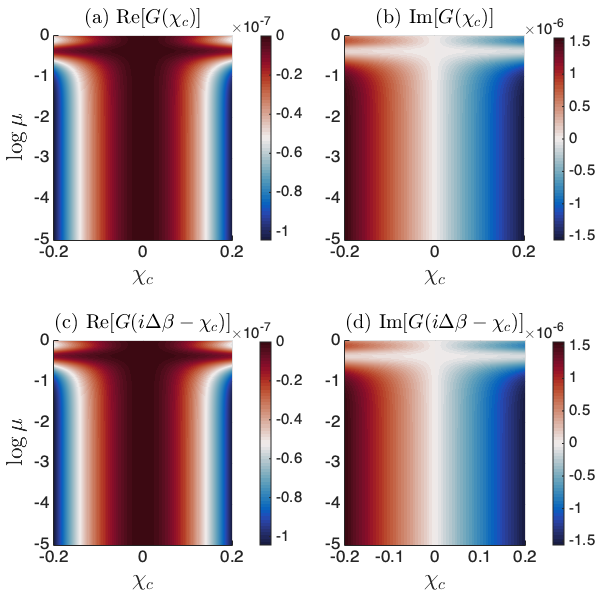}
\caption{Verifying the steady state exchange fluctuation symmetry, Eq. (\ref{eq:efs}),
for the V-shaped system using the secular $\chi$-RME Eq. (\ref{eq:secular_rme}).
(a) Real (Re) and (b) imaginary (Im) parts of $G(\chi_c)$.
(c) Real and (d) imaginary part of $G(i\Delta\beta-\chi_c)$.
Other parameters are $\epsilon_g=0$, $\epsilon_e=0.4$, $\alpha_{h,c}=0.002$, $\omega_c=50$, $T_h=0.15$ and $T_c=0.1$.}
\label{fig:efs_secular}
\end{figure}

\begin{figure}[tbh!]
 \centering
\includegraphics[width=1\columnwidth] {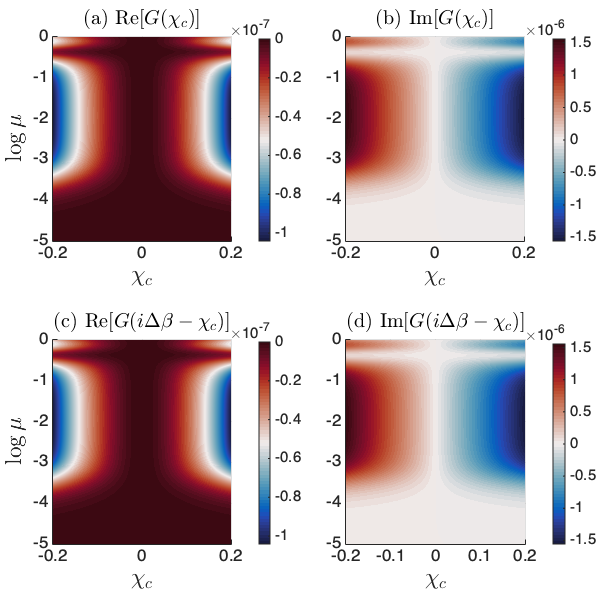}
\caption{Verifying the steady state exchange fluctuation symmetry Eq. (\ref{eq:efs})
for the V-shaped system using the full $\chi$-RME, Eq. (\ref{eq:full_rme}).
(a) Real (Re) and (b) imaginary (Im) parts of $G(\chi_c)$.
(c) Real and (d) imaginary part of $G(i\Delta\beta-\chi_c)$.
Parameters are the same as in Fig. \ref{fig:efs_secular}.}
\label{fig:efs_full}
\end{figure}

\begin{figure}[tbh!]
 \centering
\includegraphics[width=1\columnwidth] {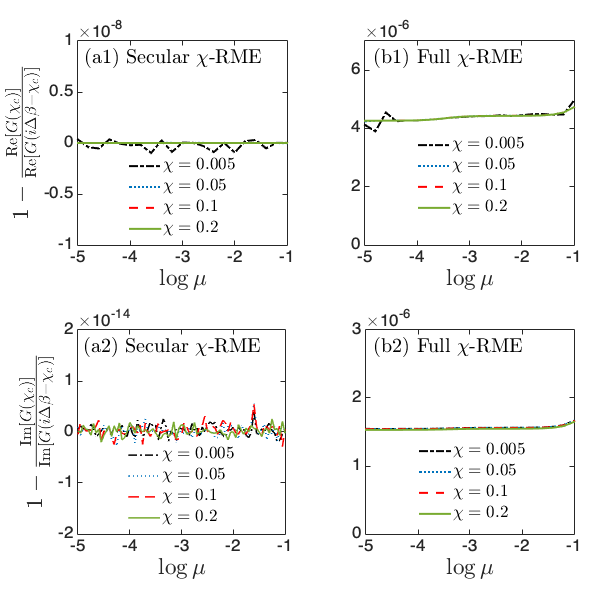}
\caption{Analysis of the SSFS in the V-shaped system using the (a) secular and (b) full Redfield equation.
We separately depict the real part of the CGF (top panels) and the imaginary part (bottom panels).
Parameters are the same as in Fig. \ref{fig:efs_secular}.}
\label{fig:ratio}
\end{figure}


To address this issue we point out the following:
(i) We are only interested in autonomous (non-driven) systems, and in their 
steady state properties. We are not aware of examples 
showing that the RME is thermodynamically improper in this case (unlike the transient regime).
(ii) Across all parameters regimes that we had analyzed here, we found that
levels population were positive (physical).
(iii) The V-shaped model [see Eq. (\ref{eq:Vmodel})] captures the behavior of coherences in the 4-level QAR, 
as discussed in Ref. \cite{Kilgour.18.PRE}.
Below, we test the fluctuation symmetry in this (two-bath) model and show that it is 
obeyed with a small numerical error.
While it is intriguing to {\it prove} that the $\chi$-RME satisfies the SSFS, our 
numerical simulations provide a strong support for this assertion, 
backing up our calculations of the current and its noise with Eq. (\ref{eq:quantity}).
(v) We tested the entropy production rate in our simulations, and found that for results presented below it was always positive.

In fact, in all the tests that we carried out in this study, which were steady state calculations, we observed that the RME lead to positive level population, positive entropy production rate, and the validity of the SSFT. 
Our simulations provide a strong numerical support for the 
thermodynamical consistency of the RME in steady state, 
even when coherences play a critical role, calling for analytic investigations.


For simplicity, we now consider a model involving just two heat baths ($v=c,~h$). 
In steady state, the input heat at the terminal $v=h$ is equal to the  heat output at $v=c$.
As such, it is sufficient to look at the fluctuation symmetry of heat exchange with a single counting parameter.
The exchange fluctuation symmetry states the following relation for the cumulant generating 
function \cite{Esposito.09.RMP,Campisi.11.RMP},
\begin{equation}\label{eq:efs}
G(\chi_c)~=~G(i\Delta\beta-\chi_c),
\end{equation}
with $\Delta\beta=1/T_h-1/T_c$. 

%

We simulate the V-shaped system coupled to two heat baths  as an example,
given that its behavior closely resembles that of Model I below [see Fig. \ref{fig:fig3} (a)] for the QAR \cite{Kilgour.18.PRE}. 
 The system Hamiltonian involves a ground ($g$) state $|g\rangle$ and two  excited ($e$) states $\{|e_1\rangle, |e_2\rangle\}$, which are coherently coupled, 
\bea
\hat H_s &=& \epsilon_g|g\rangle\langle g|+\epsilon_e\Big(|e_1\rangle\langle e_1|+|e_2\rangle\langle e_2|\Big)\nonumber\\
&&+\mu \Big(|e_1\rangle\langle e_2|+|e_2\rangle\langle e_1|\Big).
\label{eq:Vmodel}
\eea
Here $\epsilon_{g (e)}$ are the ground (excited) state energy. $\mu$ is the coupling strength between the two excited states and its magnitude can be made arbitrarily large. The transitions between ground and excited states are driven by the cold ($c$) and hot ($h$) heat baths with the following system operators associated with $\hat H_{sb}$:
\begin{equation}
\hat S_h~=~|g\rangle\langle e_1|+|e_1\rangle\langle g|,~~\hat S_c~=~|g\rangle\langle e_2|+|e_2\rangle\langle g|.
\end{equation}

First, we verify Eq. (\ref{eq:efs}) in the secular limit using Eq. (\ref{eq:secular_rme}) 
for which analytic treatments exist 
(see, for instance, Ref. \cite{Segal.18.PRE}). 
In simulations, we determine the cumulant generating function by calculating numerically the eigenvalue with the smallest real part, Eq. (\ref{eq:cgf}). 
The  dependence of $G(\chi_c)$ and $G(\Delta\beta-\chi_c)$ on the counting field $\chi_c$ and inter-state coupling strength $\mu$ is depicted in Fig. \ref{fig:efs_secular}. 
By comparing the upper and lower panels of Fig. \ref{fig:efs_secular}, it is evident that our simulations 
preserve the fluctuation symmetry Eq. (\ref{eq:efs}) in the secular limit.
Deviations between $G(\chi_c)$ and $G(\Delta\beta-\chi_c)$
were 11 orders of magnitude smaller than their value, for both real and imaginary parts 
(a quantitative analysis is included in Fig. \ref{fig:ratio}). 
Since we know that the SSFS is obeyed in the secular limit, we reason that
these deviations arise from numerical errors from the various stages involved in the procedure,
such as matrix inversion.
Particularly, we find that the real (imaginary) part of the cumulant generating function is an even (odd) function of the counting field $\chi_c$, hence Eq. (\ref{eq:quantity}) allows us to get real values for the current and its noise.

%

Next, we turn to simulations with the full $\chi$-RME Eq. (\ref{eq:full_rme}). 
The  comparison between $G(\chi_c)$ and $G(i\Delta\beta-\chi_c)$ is shown in Fig. \ref{fig:efs_full}. 
We observe that the fluctuation symmetry Eq. (\ref{eq:efs}) is preserved by the full $\chi$-RME, 
thereby suggesting the thermodynamic consistency of the method 
for evaluating the currents and fluctuations from the cumulant generating function according to Eq. (\ref{eq:quantity}). 
Interestingly, we find that the magnitude of the cumulant generating function obtained using the full Redfield dynamics is suppressed in the weak coupling regime of $\mu$ compared with the secular limit, as illustrated in Fig. \ref{fig:efs_secular}. 
We note that Ref. \cite{Kilgour.18.PRE} demonstrated current suppression in the same model arising due to the presence of finite system coherence for weak $\mu$. Our results further imply that the so-observed suppression occurs at the level of cumulant generating function and persists for finite $\chi_c$.

We interrogate the SSFS in a quantitative way in Fig. \ref{fig:ratio} by looking at the deviation of the 
ratio  $G(\chi_c)/G(i\Delta\beta-\chi_c)$ from unity, separately for the real and imaginary parts.
The errors in the secular $\chi$-RME are certainly numerical since the SSFS is obeyed in this case.
Turning to the full $\chi$-RME, we note higher deviations from unity 
for both real and imaginary parts compared to the secular case.
However, deviations from perfect symmetry do not depend on the intersite coupling $\mu$ or the counting parameter $\chi$.
Particularly for $\mu$, coherence effects show up in this model for  $\log \mu \lesssim-3.5$ \cite{Kilgour.18.PRE}. 
The fact that the agreement between $G(\chi_c)$ and $G(i\Delta\beta-\chi_c)$ does not depend on $\mu$
suggests that the error is accumulated by numerical operations rather than reflecting a fundamental violation.

Note that the secular  $\chi$-RME calculation on the V-shaped model is performed by studying the eigenvalue of a $3\times3$ matrix. 
In contrast, the full $\chi$-RME calculation
is performed by diagonalizing a $9\times9$ matrix. As such, the computational effort, and thus error accumulation 
in these two cases is quite different.

To the best of our knowledge, simulations in Figs. \ref{fig:efs_full}-\ref{fig:ratio} are the first strong numerical indication
of the validity of the SSFS in the full Redfield formalism for multi-level systems, 
and for arbitrarily large $\chi$.
We highlight that in fact we found that all eigenvalues of 
$\mathbb{L}_{\chi}$ obey the SSFS symmetry. %
Finally, predictions from the $\chi$-RME are obviously not exact 
given the perturbation approximation involved. 
In validating the SSFS  for the $\chi$-RME
we point out that  though inaccurate,
calculations of high order cumulants are physical (thermodynamically consistent).


\section{Quantum absorption refrigerators: Results}
\label{sec:3}

\begin{figure}[tbh!]
 \centering
\includegraphics[width=0.8\columnwidth] {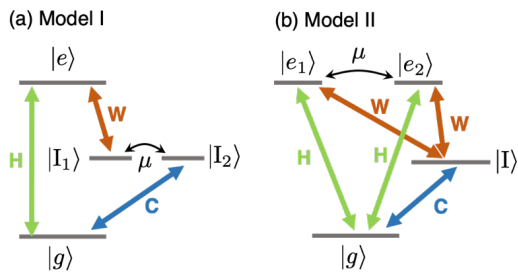} 
\caption{Level schemes of the working medium of two quantum absorption refrigerators considered in this study. 
Transitions in green, red and blue are triggered by hot (H), work (W) and cold (C) heat baths, respectively. $g,~e$ and I denote the ground, excited and intermediate states in the local site basis, respectively. 
$\mu$ denotes the coupling strength between degenerate states (in the local basis). 
Model I was recently studied in Ref. \cite{Kilgour.18.PRE}. 
Model II with $\mu=0$ was investigated in Ref. \cite{Holubec.18.JLTP}.} 
\label{fig:fig3}
\end{figure}

In Sec. \ref{sec:2S}, we established the validity of the nonsecular 
$\chi$-RME, Eq. (\ref{eq:full_rme}). Equipped with this method,
we now study the steady-state behavior of QARs and contrast it to the secular limit, Eq. (\ref{eq:secular_rme}). 
We consider two distinct four-level QARs, with their level diagrams depicted in Fig. \ref{fig:fig3}. 
To operate as a QAR, three heat baths, $v=h,c,w$ are included; the QAR continuously pumps heat from the cold ($c$) bath to the hot ($h$) bath consuming power from the work ($w$) bath. 

We follow the standard setting, that QARs are composed of several 
subsystems that are spatially separated \cite{Mitchison.19.CP}.  
This allows us to consider a selective coupling scheme, namely, each transition 
is triggered by an individual heat bath. 
With this prerequisite, Ref. \cite{Kilgour.18.PRE} has shown that system coherences have deleterious
 effects on the cooling power of Model I [Fig. \ref{fig:fig3} (a)]. 
On the contrary, Ref. \cite{Holubec.18.JLTP} found that system coherences can boost the cooling power of QARs 
in Model II [Fig. \ref{fig:fig3} (b)] with $\mu=0$. 
Nevertheless, both studies (cf. Refs. \cite{Kilgour.18.PRE,Holubec.18.JLTP}) were focused on the cooling power without examining its fluctuation behaviors.  
A recent study \cite{Holubec.19.JCP} had addressed the fluctuation behavior of Model II with $\mu=0$, that is, with an eigenenergy degeneracy.
Close to maximum cooling current, this model displays a special behavior, with a non-unique steady state.
Here, we consider the non-degenerate scenario with $\mu\neq0$, which  is very different, always resulting in a unique steady state solution. %

Below, we perform a thorough investigation of Models I and II with a focus on the interplay 
of system coherences and power fluctuations. 
In what follows, we use the subscript `FR' to denote results from the 
full $\chi$-RME Eq. (\ref{eq:full_rme}) and `S' for results from the secular $\chi$-RME Eq. 
(\ref{eq:secular_rme}).

\begin{figure*}[tbh!]
 \centering
\includegraphics[width=2\columnwidth] {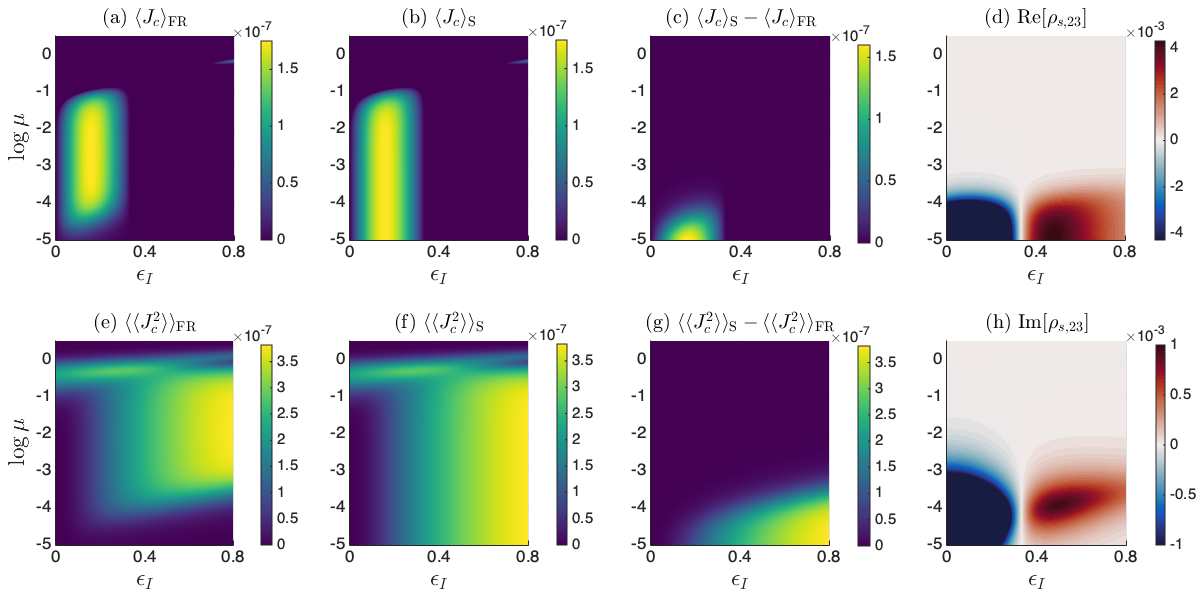} 
\caption{
Current, noise and coherences in Model I.
(a) Cooling power $\langle J_c\rangle_{\mathrm{FR}}$ 
obtained from the  full $\chi$-RME [Eq. (\ref{eq:full_rme})].
(b) Cooling power $\langle J_c\rangle_{\mathrm{S}}$ obtained 
(b) from the secular $\chi$-RME  [Eq. (\ref{eq:secular_rme})]. 
The deep blue background in  (a) and (b) marks the no-cooling region with $\langle J_c\rangle<0$ (not shown). 
(c) Power difference $\langle J_c\rangle_{\mathrm{S}}-\langle J_c\rangle_{\mathrm{FR}}$ in the cooling region. 
(e) Current fluctuations $\langle\langle J_c^2\rangle\rangle_{\mathrm{FR}}$ in the full $\chi$-RME 
[Eq. (\ref{eq:full_rme})]. (f) Current fluctuations $\langle\langle J_c^2\rangle\rangle_{\mathrm{S}}$ in the
 secular $\chi$-RME [Eq. (\ref{eq:secular_rme})]. 
(g) Difference in fluctuations between the secular and non-secular calculations. (d) and (h) show the real (Re) and imaginary (Im) parts of the off-diagonal element of 
the reduced steady state density matrix, $\rho_{s,23}$, obtained from the full Redfield master equation. 
 Parameters are $\alpha_{w,h,c}=0.002$, $\omega_c=50$, $T_w=0.2$, $T_h=0.15$, $T_c=0.1$, $\epsilon_e=1$.} 
\label{fig:M1_1}
\end{figure*}
%
\begin{figure*}[tbh!]
\centering
\includegraphics[width=2\columnwidth]{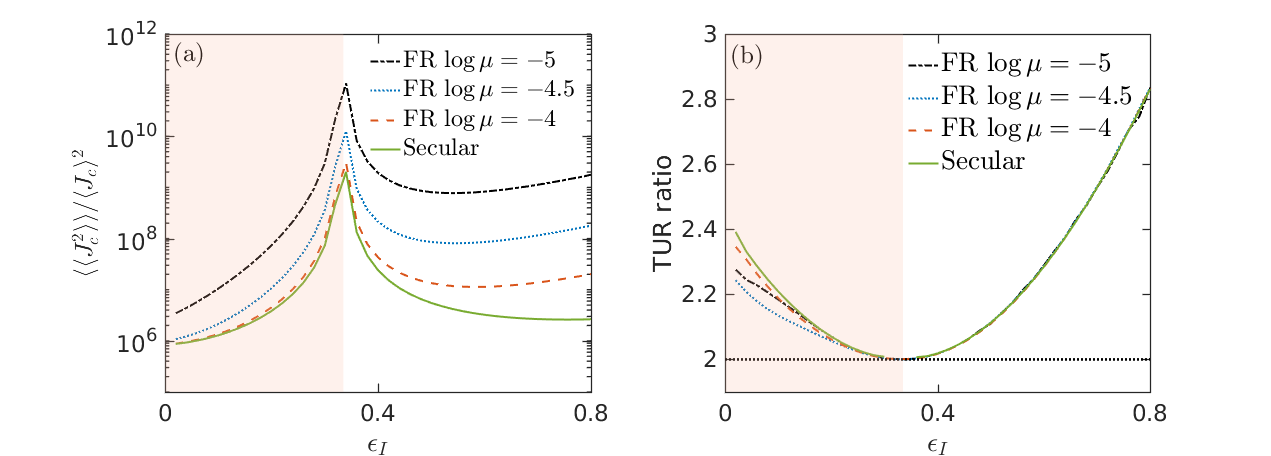}  
\caption{
Relative noise and the TUR in Model I.
(a) Relative noise $\langle\langle J_c^2\rangle\rangle/\langle J_c\rangle^2$ for different $\log\mu$. 
using the full $\chi$-RME [Eq. (\ref{eq:full_rme})].
The secular limit [Eq. (\ref{eq:secular_rme})] does not depend on $\mu$.
(b) Corresponding 
TUR ratio $\langle\langle J_c^2\rangle\rangle \langle \sigma\rangle/(\langle J_c\rangle^2)$.  
%
The TUR bound (at two) is highlighted by a horizontal black dotted line. 
The shaded region marks the cooling region. 
Other parameters are the same with Fig. \ref{fig:M1_1}.} 
\label{fig:M1_2}
\end{figure*}

\subsection{Model I}
In the local site basis, the working medium of Model I (see Fig. \ref{fig:fig3} (a)) 
is described by a Hamiltonian
\bea
\hat H_s^I &=& \epsilon_g|g\rangle\langle g|+\epsilon_e|e\rangle\langle e|\nonumber\\
&&+\epsilon_I(|\mathrm{I}_1\rangle\langle \mathrm{I}_1|+|\mathrm{I}_2\rangle\langle \mathrm{I}_2|)+\mu(|\mathrm{I}_1\rangle\langle \mathrm{I}_2|+|\mathrm{I}_2\rangle\langle \mathrm{I}_1|).
\eea
The system includes a ground (g) state $|g\rangle$, an excited (e) 
state $|e\rangle$, and two degenerate intermediate (I) levels 
($|\mathrm{I}_1\rangle,~|\mathrm{I}_2\rangle$) connected by a coherent hoping rate $\mu$.  Note that the levels are degenerate in the local basis,
and non-degenerate in the global basis.
We set the reference energy at $\epsilon_g=0$. 
The system's operators involved in the system-bath interaction $\hat H_{sb}$ have the forms
\bea
\hat S_c &=& |g\rangle\langle \mathrm{I}_2|+|\mathrm{I}_2\rangle\langle g|,\nonumber\\
\hat S_h &=& |g\rangle\langle e|+|e\rangle\langle g|,\nonumber\\
\hat S_w &=& |\mathrm{I}_1\rangle\langle e|+|e\rangle\langle \mathrm{I}_1|.
\eea
After ordering the labels of eigenstates of $\hat H_s$ such that 
$\hat H_s=\sum_{i=1}^4E_i|i\rangle\langle i|$ with $E_i<E_{i+1}$, we rewrite the above system operators in the energy basis
\bea
\hat S_c &=& \frac{1}{\sqrt{2}}\left(|1\rangle\langle 3|-|1\rangle\langle 2|+\mathrm{H.c.}\right),\nonumber\\
\hat S_h &=& |1\rangle\langle 4|+|4\rangle\langle 1|,\nonumber\\
\hat S_w &=& \frac{1}{\sqrt{2}}\left(|2\rangle\langle 4|+|3\rangle\langle 4|+\mathrm{H.c.}\right).
\eea
%

\begin{figure}[tbh!]
 \centering
\includegraphics[width=1\columnwidth] {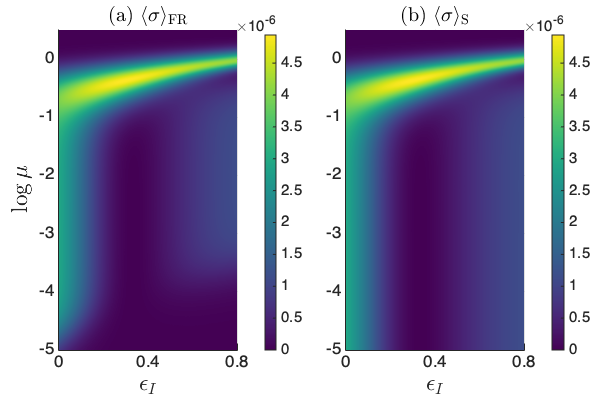}
\caption{
Verification of the positivity of the  entropy production rate for Model I
for the (a) full Redfield case, (b) secular limit.
Parameters are the same as in Fig. \ref{fig:M1_1}.}
\label{fig:entropy}
\end{figure}


In Fig. \ref{fig:M1_1} we display simulation results for the
current $\langle J_c\rangle$ (when referring to as cooling power we implicitly imply $\langle J_c\rangle>0$) 
and its noise $\langle\langle J_c^2\rangle\rangle$. 
From panels \ref{fig:M1_1} (c) and (d) or (h) it is evident that finite system coherences, 
as quantified by the real or imaginary part of the steady state density matrix element $\rho_{s,23}$ is responsible for the suppression of the cooling power in the weak $\mu$ regime relative to the secular limit 
\cite{Kilgour.18.PRE}. 
Note that the deep blue background in panels (a) and (b) marks the no-cooling region, with $\langle J_c\rangle<0$.
In other words, for presentation purposes we present the no-cooling region as zero cooling currents.
The cooling region is marked in Fig. \ref{fig:M1_2}, where it appears when $\epsilon_I\lesssim0.332$. 
Note that for the secular limit, results in all figures here and below were independent of $\mu$ for the small $\mu$ considered.  

In Fig. \ref{fig:M1_1} (e)-(f), we further show the current fluctuations. 
We find from panel \ref{fig:M1_1} (e) and (f) that fluctuations become pronounced in the non-cooling region. 
Interestingly, the system coherence  suppresses the current fluctuation but {\it in the non-cooling region}, as can be seen from the comparison between Fig. \ref{fig:M1_1} (e) and (f). 
Noting that the signs of system coherence in the cooling and non-cooling regions are opposite
[see Fig. \ref{fig:M1_1} (d) and (h)], it is then clear that in Model I negative coherence in the cooling region 
induces power suppression, while positive coherence in the non-cooling region is responsible for the suppression
of current fluctuations. 
Nevertheless, as we are interested in the cooling region, such a fluctuation suppression is of no practical use. 
Intriguingly, the marginal regions between finite coherences in Fig. \ref{fig:M1_1} (d) and (h) marks the boundary of cooling and non-cooling regions, namely, the transition from cooling to non-cooling regions is associated with a sign change of system coherence.

We now look at the relative noise,
 $\langle\langle J_c^2\rangle\rangle/\langle J_c\rangle^2$ and the TUR 
ratio $\langle\langle J_c^2\rangle\rangle \langle \sigma\rangle/\langle J_c\rangle^2$,
see Fig. \ref{fig:M1_2}. 
In panel \ref{fig:M1_2} (a) we observe that the relative noise obtained from the 
full $\chi$-QME is greatly enhanced due to the presence of coherences 
in both the cooling and non-cooling regions,  compared to the secular limit. 
Ideally, the relative noise tends to infinity at the exact boundary of cooling and non-cooling regions as $\langle J_c\rangle=0$, however, we can only see finite peak structures centered around the boundary from panel \ref{fig:M1_2} (a) since we utilize discretized values for $\epsilon_I$ in simulations and may not be able to reach the exact boundary. 

The TUR ratio is plotted in panel \ref{fig:M1_2} (b). We observe that: 
(i) The TUR ratio is always above the classical bound 
set by the TUR Eq. (\ref{eq:tur}). 
(ii) The TUR ratio saturates to the bound at the point when the entropy production is exactly zero
and the refrigerator crosses into the no-cooling region.
(iii) Coherences slightly reduce the TUR ratio, relative to the secular limit, yet the behavior is non-monotonic.
Nevertheless, coherences only mildly reduce the TUR ratio from the secular limit.
The calculation of the TUR ratio at the cooling-no-cooling boundary region is nontrivial numerically.
This is due to the nontrivial cancellation between entropy production and the relative noise taking place when approaching the boundary. 
As such, very close to the boundary region the TUR ratio was not evaluated. 

Back to the fundamental question as to whether the non-secular Redfield equation is suitable
 (thermodynamically consistent) for studying current noise.
In  Fig. \ref{fig:entropy} we verify that the entropy production rate 
is always positive with our parameters. In fact, we have not observed negative entropy production rates
in any of our simulations.
As such, we hypothesize that the Redfield equation is thermodynamically consistent  in the steady state limit
with $\langle \sigma\rangle >0$, beyond our case study.
This result does not exclude the possibility of observing fundamental deficiencies with the Redfield equation
in the transient regime.

Concluding this Section: The TUR, Eq. (\ref{eq:tur}), is satisfied by Model I in the presence of system coherences. 
Notably, the TUR ratio of Model I is almost independent of the inter-site 
coupling strength $\mu$, and it almost coincides with that of the secular limit, thereby implying that the 
coherence-induced enhancement of the relative noise is compensated by the coherence-induced reduction 
of entropy production rate, which is proportional to the heat currents.

%
\begin{figure*}[tbh!]
 \centering
\includegraphics[width=2\columnwidth] {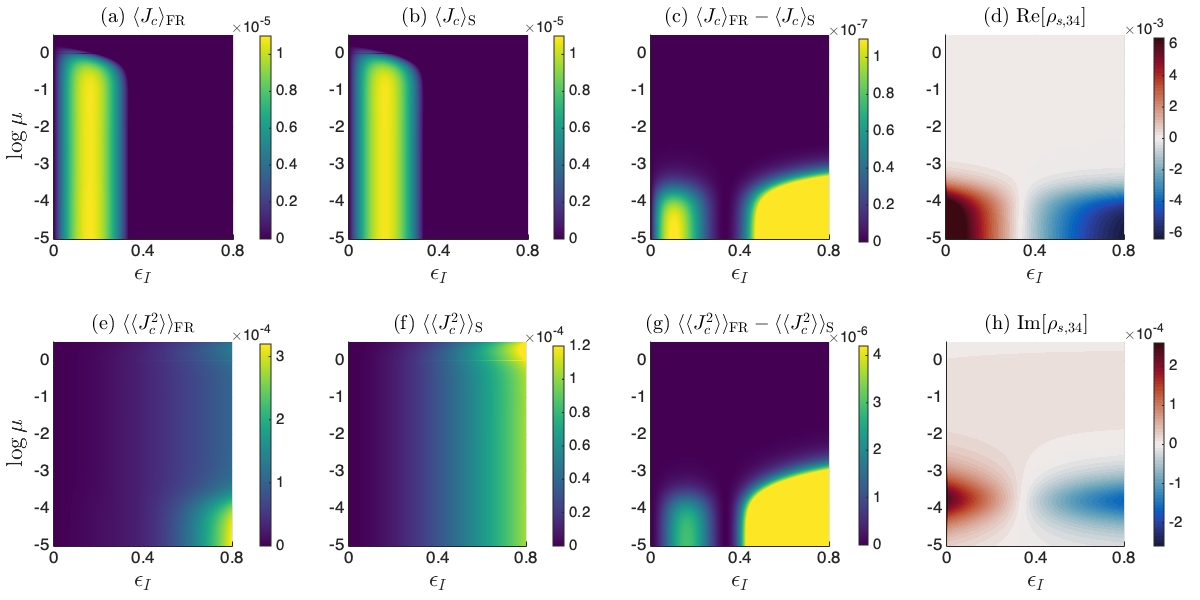} 
\caption{Current, fluctuations and coherence in Model II.
(a) Cooling power $\langle J_c\rangle_{\mathrm{FR}}$ obtained from the full $\chi$-RME 
[Eq. (\ref{eq:full_rme})]. 
(b) Cooling power $\langle J_c\rangle_{\mathrm{S}}$ from the secular $\chi$-RME [Eq. (\ref{eq:secular_rme})]. 
The deep blue backgrounds in (a) and (b) mark the no-cooling region with $\langle J_c\rangle<0$ (not shown). 
(c) The difference  in the cooling current, 
$\langle J_c\rangle_{\mathrm{FR}}-\langle J_c\rangle_{\mathrm{S}}$ in the whole region. 
(e) Current fluctuations $\langle\langle J_c^2\rangle\rangle_{\mathrm{FR}}$ from the full 
$\chi$-RME [Eq. (\ref{eq:full_rme})]. 
(f) Current fluctuations $\langle\langle J_c^2\rangle\rangle_{\mathrm{S}}$ in the secular $\chi$-RME [Eq. (\ref{eq:secular_rme})].
(g) The difference $\langle\langle J_c^2\rangle\rangle_{\mathrm{FR}}-\langle\langle J_c^2\rangle\rangle_{\mathrm{S}}$ in the whole region. (d) and (h) show the real (Re) and imaginary (Im) parts of the off-diagonal element of 
the reduced steady state density matrix, $\rho_{s,34}$, obtained from the full Redfield master equation.
 Parameters are $\alpha_c=0.002$, $\alpha_{h1}=0.8\alpha_c$, $\alpha_{h2}=\alpha_c$, $\alpha_{w1}=\alpha_c$, $\alpha_{w2}=\alpha_c$, $\omega_c=50$, $T_w=2$, $T_h=0.6$, $T_c=0.25$, $\epsilon_e=1$.} 
\label{fig:M2_1}
\end{figure*}
%
\begin{figure*}[tbh!]
 \centering
\includegraphics[width=2\columnwidth] {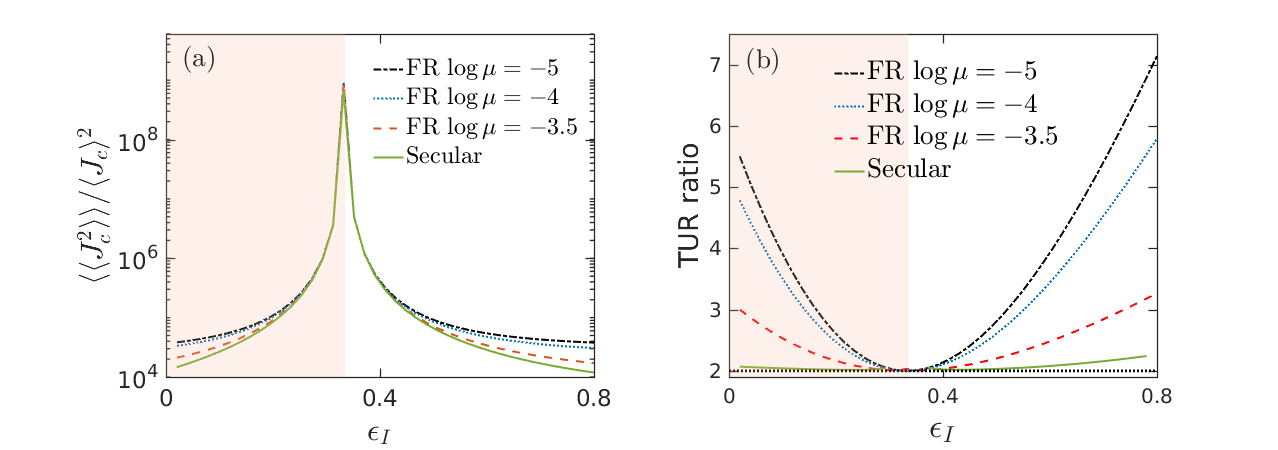} 
\caption{
Relative noise and the TUR in Model II.
(a) Relative noise $\langle\langle J_c^2\rangle\rangle/\langle J_c\rangle^2$ for different $\log\mu$.
using the full $\chi$-RME [Eq. (\ref{eq:full_rme})].
The secular limit [Eq. (\ref{eq:secular_rme})] does not depend on $\mu$.
(b) Corresponding
TUR ratio $\langle\langle J_c^2\rangle\rangle \langle \sigma\rangle/(\langle J_c\rangle^2)$.  
%
The TUR bound (at two) is highlighted by a horizontal black dotted line.
The shaded region marks the cooling region.
Other parameters are the same with Fig. \ref{fig:M2_1}.}
\label{fig:M2_3}
\end{figure*}
%
\subsection{Model II}
We turn to Model II for a QAR, as illustrated in Fig. \ref{fig:fig3} (b). 
Previously, Refs. \cite{Holubec.18.JLTP,Holubec.19.JCP} demonstrated that this model 
can have a coherence-induced enhancement of the cooling power; below we confirm this scenario and further 
show in the Appendix  
that this model also 
permits an adverse effect of system coherence on the cooling power, 
when varying the system-bath coupling strengths.

The working medium of Model II is described by a four level Hamiltonian in the local basis
\bea
\hat H_s^{II} &=& \epsilon_e(|e_1\rangle\langle e_1|+|e_2\rangle\langle e_2|)+\mu(|e_1\rangle\langle e_2|+|e_2\rangle\langle e_1|)\nonumber\\
&&+ \epsilon_g|g\rangle\langle g|+\epsilon_I|\mathrm{I}\rangle\langle \mathrm{I}|.
\eea
The system involves a ground (g) state $|g\rangle$, an intermediate (I) state $|\mathrm{I}\rangle$ and two degenerate excited (e) states ($|e_1\rangle,~|e_2\rangle$) that are 
connected by a coherent hoping rate, $\mu$. 
We set the reference energy at $\epsilon_g=0$. 
After ordering the labels of eigenstates of $\hat H_s$ such that $\hat H_s=\sum_{i=1}^4E_i|i\rangle\langle i|$ with $E_i<E_{i+1}$, 
we get the following system operators, involved in the system-bath interaction $\hat H_{sb}$,
 in the energy basis \cite{Holubec.18.JLTP}
\bea
\hat S_c &=&|1\rangle\langle 2|+|2\rangle\langle 1|,\nonumber\\
\hat S_{h} &=& \sqrt{\alpha_{h2}}|1\rangle\langle 4|+\sqrt{\alpha_{h1}}|1\rangle\langle 3|+\mathrm{H.c.},\nonumber\\
\hat S_{w} &=& \sqrt{\alpha_{w2}}|2\rangle\langle 4|+\sqrt{\alpha_{w1}}|2\rangle\langle 3|+\mathrm{H.c.}.
\eea
The real-valued parameters $\alpha_{vk}$ are dimensionless; they are taken from the spectral density function
$\gamma_v(\omega)$ in Eq. (\ref{eq:RR}); recall that we model the spectral density function as
$\gamma_{v}=\alpha_v \omega e^{-\omega/\omega_c}$. 
Here for convenience we absorbed $\alpha_{vk}$ into the definitions of $\hat S_{h,w}$  so as to 
allow scenarios where different transitions 
induced by the same bath are enhanced by different coupling strengths.


Fig. \ref{fig:M2_1} depicts an example where system coherences {\it boost} the cooling power, 
as predicted in Refs. \cite{Holubec.18.JLTP,Holubec.19.JCP}. 
This is highlighted in 
panel \ref{fig:M2_1} (c), where we show the current difference 
$\langle J_c\rangle_{\mathrm{FR}}-\langle J_c\rangle_{\mathrm{S}}$ in both the cooling and no-cooling regions. 
As can be seen, the cooling power $\langle J_c\rangle_{\mathrm{FR}}$ in the cooling region is slightly enhanced 
compared with $\langle J_c\rangle_{\mathrm{S}}$ when $\mu$ is relatively small. 
We attribute this cooling power enhancement to the finite positive system coherence in that region 
as indicated in Fig. \ref{fig:M2_1} (d) and (h). However,  panel \ref{fig:M2_1} (g) shows that 
this cooling power enhancement comes at the price of an enhanced power fluctuation. 
%
%
%
%
Similarly to Model I, here we also find that the marginal region between nonzero coherences (Fig. \ref{fig:M2_1} (d) and (h)) marks the boundary  between the cooling and the no-cooling regions.

While in Fig. \ref{fig:M2_1}  system coherences enhance the cooling current,
in the  Appendix  we show the opposite effect within the same model, but using
a different value for $\alpha_{h1}$.
This coherence-induced suppression of the cooling power (similarly to Model I)
 becomes evident by combining the information of Fig. \ref{fig:a_1} (c) and (d) or (h).
 
Even though system coherences can either suppress or enhance the cooling power 
in Model II (depending on the system-bath coupling strengths), 
in Fig. \ref{fig:M2_3} (a) we again observe that the relative noise 
obtained from the full $\chi$-RME is larger than that obtained from the secular $\chi$-RME. 
Furthermore, the relative noise approaches the secular limit near the boundary between the cooling and 
no-cooling regions. 
Nevertheless, from Fig. \ref{fig:M2_3} (b)  (together with Fig. \ref{fig:a_2} (b) in the Appendix) 
we find that the TUR holds as well in Model II.
Interestingly, the TUR ratio is enhanced by coherences relative to the secular limit:
In Fig. \ref{fig:M2_3} (b), we see that the TUR ratio in the secular case is very close to the bound within the whole range of $\epsilon_I$, while it is factor of 3 greater in the coherent case.
%

Overall, we found that in model II the effect of coherences on the cooling current was very mild; we did not perform detailed simulations to identify region of more substantial cooling effect as this was not the objective of this work. Our main conclusion, which holds for all cases examined here is that while coherences may boost or suppress the cooling current, their effect on fluctuations is adverse, thus validating the standard TUR.


\section{Summary}\label{sec:4}

In summary, we addressed the interplay of system coherences and fluctuations of the cooling current
in the performance of steady-state QARs. 
Using a Redfield master equation with a full counting statistics information, 
we obtained the behaviors of the cooling power and its fluctuations, 
for steady-state QAR models with (or without) system coherences. 
Remarkably, we found that the relative noise of the cooling power was always enhanced 
in the presence of system coherence, even though the cooling power itself was either suppressed or enhanced, 
depending on the model and its parameters.
As a result, we confirmed that the TUR derived for classical Markov-jump processes holds 
for steady-state QARs in the presence of system coherence;
the performance of the steady-state QARs is still constrained by the classical tradeoff relation.

Our results apply to scenarios where the Redfield master equation can be justified. 
Although a general proof within the framework of the Redfield master equation 
is still missing,  we expect that our results are general:
System coherence corroborates the classical TUR.
After all, system coherences correspond to additional quantum fluctuations, on top of thermal ones to the QTMs. 
In fact, a recent study on steady-state quantum heat engines \cite{Bret.20.A} reached a similar conclusion 
on the role of system coherences in validating the TUR. 
However, if cyclic instead of steady-state QTMs are concerned, 
our conclusions need to be revisited as a recent study \cite{Cangemi.20.PRB} suggested that system coherences 
can help to violate a TUR specific for periodic-driven systems.

Is there a ``quantum" advantage for thermal machines, compared to their classical counterparts? 
While the power output may be enhanced due to coherences---depending on the model employed,
here we point out to what seems to be a more general adverse effect of quantum coherences: 
According to our examples, QTMs suffer more pronounced thermodynamic fluctuations 
arising due to finite system coherence, compared to the incoherent analogue.
Our findings further imply that fluctuations should be considered when assessing whether the
system coherence is a useful resource to the operation of QTMs.

Altogether, our contributions are: 
(i) We verified with simulations that the counting-field dressed Redfield master 
equation satisfies the SSFS for heat transfer.
(ii) We demonstrated that system coherences intensify relative current fluctuations, 
irrespective of the impact on the cooling power.
(iii) We showed that the classical TUR holds in the presence of system coherences.

As a final remark, our study indicates that the standard (classical) TUR for steady state transport
is valid in the weak system-bath coupling regime (see also Ref. \cite{Saryal}).
To observe violations---thus circumvent the classical tradeoff relation for thermal machines---one should turn to the nonperturbative system-bath coupling regime \cite{BijayNJPSB}, 
where system-bath entanglement and nonmarkovianity play a decisive role in breaking the TUR \cite{Saryal}.

\begin{acknowledgments}
The authors acknowledge support from the Natural Sciences and Engineering 
Research Council (NSERC) of Canada Discovery Grant
and the Canada Research Chairs Program.
\end{acknowledgments}

\appendix
\renewcommand{\theequation}{A\arabic{equation}}
\renewcommand{\thefigure}{A\arabic{figure}}
\setcounter{equation}{0}  
\setcounter{figure}{0}
\section{Additional simulations  for Model II}\label{a:1}
\begin{figure*}[tbh!]
 \centering
\includegraphics[width=2\columnwidth] {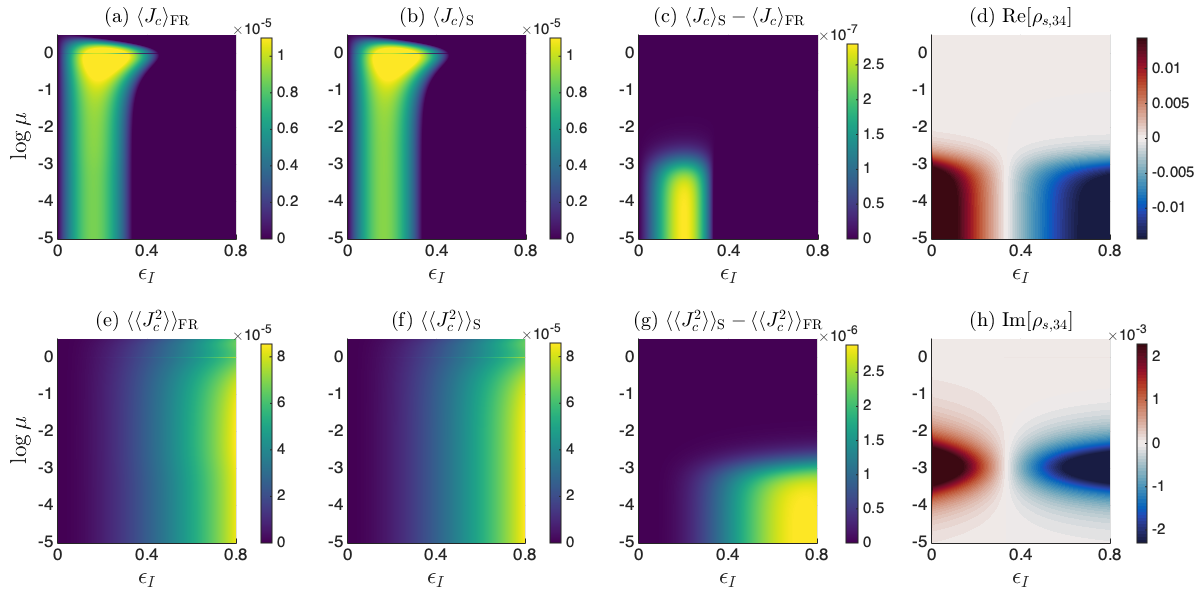} 
\caption{
Current, fluctuations and coherence in Model II with $\alpha_{h1}=0.1\alpha_c$; 
other parameters are the same as in Fig. \ref{fig:M2_1}.
(a) Cooling power $\langle J_c\rangle_{\mathrm{FR}}$ obtained from the full $\chi$-RME [Eq. (\ref{eq:full_rme})].
 (b) Cooling power $\langle J_c\rangle_{\mathrm{S}}$ from the secular $\chi$-RME [Eq. (\ref{eq:secular_rme})]. 
The deep blue backgrounds in (a) and (b) mark the no-cooling region with $\langle J_c\rangle<0$ (not shown). 
(c) Difference in cooling current $\langle J_c\rangle_{\mathrm{S}}-\langle J_c\rangle_{\mathrm{FR}}$ in the whole region. (e) Current fluctuations $\langle\langle J_c^2\rangle\rangle_{\mathrm{FR}}$ from the full 
$\chi$-RME [Eq. (\ref{eq:full_rme})]. 
(f) Current fluctuations $\langle\langle J_c^2\rangle\rangle_{\mathrm{S}}$ in the secular $\chi$-RME [Eq. (\ref{eq:secular_rme})]. 
(g) The difference $\langle\langle J_c^2\rangle\rangle_{\mathrm{S}}-\langle\langle J_c^2\rangle\rangle_{\mathrm{FR}}$ in the whole region.  (d) and (h) show the real (Re) and imaginary (Im) parts of the off-diagonal element of 
the reduced steady state density matrix, $\rho_{s,34}$, obtained from the full Redfield master equation.
} 
\label{fig:a_1}
\end{figure*}

\begin{figure*}[tbh!]
 \centering
\includegraphics[width=1.8\columnwidth] {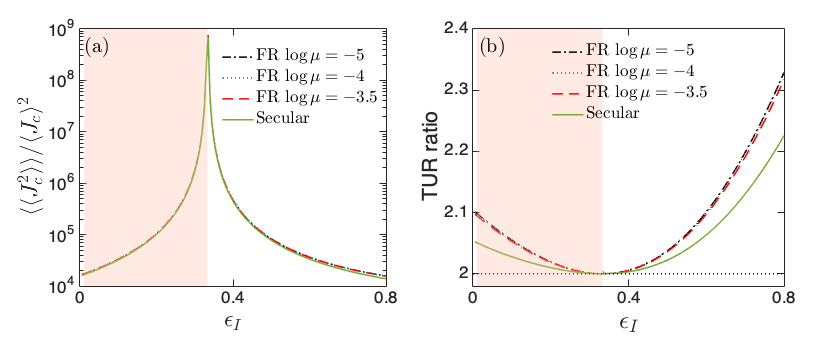} 
\caption{
Relative noise and the TUR in Model II with $\alpha_{h1}=0.1\alpha_c$; other parameters are the same with Fig. \ref{fig:M2_1}.
(a) Relative noise $\langle\langle J_c^2\rangle\rangle/\langle J_c\rangle^2$ for different $\log\mu$.
using the full $\chi$-RME [Eq. (\ref{eq:full_rme})].
The secular limit [Eq. (\ref{eq:secular_rme})] does not depend on $\mu$.
(b) Corresponding
TUR ratio $\langle\langle J_c^2\rangle\rangle \langle \sigma\rangle/(\langle J_c\rangle^2)$.  
%
The TUR bound (at two) is highlighted by a horizontal black dotted line.
The shaded region marks the cooling region. 
}
\label{fig:a_2}
\end{figure*}

In this appendix, we show that Model II can also allow for a coherence-induced suppression of the  cooling power when varying the system-bath coupling strength, say, $\alpha_{h1}$. A representative set of results with $\alpha_{h1}=0.1\alpha_c$ is depicted in Figs. \ref{fig:a_1} and \ref{fig:a_2}. The cooling power suppression becomes evident by inspecting Fig. \ref{fig:a_1} (c) and (d) or (h). 
%

%
\end{document}